\documentclass{sig-alternate}
\usepackage{algorithm}
\usepackage{algpseudocode}
\usepackage{subfigure}
\usepackage{hyperref}
\usepackage{times} 

\usepackage{enumitem}
\setlist{nolistsep,leftmargin=*}

\newtheorem{proposition}{Proposition}
\newtheorem{theorem}{Theorem}
\newtheorem{definition}{Definition}
\newfloat{algo}{t}{lop}
\floatname{algo}{Algorithm}

\makeatletter
\let\@copyrightspace\relax
\makeatother

\begin{document}

\title{Privacy Preserving Record Linkage via {\ttlit grams} Projections}

%
%
%
%
%

\numberofauthors{2} 
%
\author{
%
%
\alignauthor
Luca Bonomi\\
       \affaddr{Emory University}\\
       \affaddr{Atlanta, USA}\\
       \email{lbonomi@mathcs.emory.edu}
\alignauthor
Li Xiong\\
       \affaddr{Emory University}\\
       \affaddr{Atlanta, USA}\\
       \email{lxiong@mathcs.emory.edu}
\and
\alignauthor
Rui Chen\\
	\affaddr{Concordia University}\\
	\affaddr{Montreal, Canada}\\
	\email{ru\_che@encs.concordia.ca}
\alignauthor
Benjamin C. M. Fung\\
     	\affaddr{Concordia University}\\
    	\affaddr{Montreal, Canada}\\
       \email{fung@ciise.concordia.ca}
}

\maketitle

\begin{abstract}
Record linkage has been extensively used in various data mining applications involving sharing data. While the amount of available data is growing, the concern of disclosing sensitive information poses the problem of utility vs privacy. In this paper, we study the problem of private record linkage via secure data transformations. In contrast to the existing techniques in this area, we propose a novel approach that provides strong privacy guarantees under the formal framework of differential privacy. We develop an embedding strategy based on frequent variable length grams mined in a private way from the original data.
We also introduce personalized threshold for matching individual records in the embedded space which achieves better linkage accuracy than the existing global threshold approach. Compared with the state-of-the-art secure matching schema \cite{Scannapieco:2007}, our approach provides formal, provable privacy guarantees and achieves better scalability while providing comparable utility.

\end{abstract}

\section{Introduction}

Record linkage \cite{winkler06overview,DBLP:journals/tkde/ElmagarmidIV07} plays a central role in many data integration and data mining tasks that involve data from multiple sources.  It is the process of identifying records that refer to the same real world entity across different sources. It is extensively used in many applications, for example, in linking medical data of the same patient across different hospitals in the country or in collecting the credit history of users from several sources.  However, many of these data may contain sensitive personal information that could disclose individual privacy. For this reason, the problem of privacy preserving record linkage has drawn considerable attention over recent years.  The objective is to allow two parties to identify records that are close to each other according to some distance function, such that no additional information about the data records other than the result is disclosed to any party.

In the existing literature, several privacy models and techniques have been proposed, and they can be mainly categorized into a few categories: secure transformation \cite{Al-Lawati:2005,Blind,Scannapieco:2007,Bloom}, Secure Multiparty Computation (SMC) \cite{crypto:2008,Yao:86} and hybrid methods \cite{EvangelistaCSM10,Inan:2008,Inan:2010}. While SMC techniques guarantee the privacy and the security by using cryptographic algorithms, it is computationally prohibitive in practice. On the other hand, secure transformation techniques achieve the privacy and security using data transformations which lead to faster algorithms. However, they have some limitations in the privacy model used, for example $k$-anonymization \cite{Sweeney:2002}, where high levels of protection typically implies a great loss of accuracy in the final results. Finally, hybrid techniques combine the previous two strategies leading to a good trade-off between privacy and utility. As a drawback, many of these approaches involve heuristics to reduce the computational cost and they suffer on the privacy model as the secure transformation techniques.  

In this paper we present a new secure data transformation method  based on an embedding technique. In addition, we provide formal guarantees for the individual user privacy by developing a secure mechanism under the framework of  \textit{differential privacy} \cite{Dwork06}. Our approach presents interesting new features such as the use of differential privacy techniques in constructing  a base for embedding the original data that gives a better representation of the data with respect to a random base. As a proof-of-concept, we focus on string records in this paper, and perform approximate matching of the records based on a similarity criterion. 

{\flushleft{\bf Our contributions:}} 
\begin{itemize}
\item We propose a novel embedding strategy based on frequent variable length grams to map string records into vectors in the real space. We show that the use of frequent variable length grams substantially increases the utility of the results with respect to random strings.
\item The private record linkage protocol proposed satisfies the differential privacy framework which provides formal guarantees of individual privacy. In addition, this allows our strategy to have a trade-off between privacy level and utility that the user can choose by changing the private parameter $\epsilon$. 
\item Our strategy allows approximate matching, and in this framework we introduced the concept of \textit{personalized threshold} for matching the strings embedded in the new space. Contrary to matching approaches that use a global threshold, our approach better captures the characteristics of each string using a personalized threshold for each record. Indeed, a  global threshold does not take advantage of the similarity of particular records. In addition, a global threshold mechanism generally requires a prior knowledge to determine the optimal threshold. On the other hand, our strategy automatically computes a personalized threshold for each string by exploiting the data characteristics. 
\item Finally, we present a set of empirical experiments using real world datasets showing the benefit of our approach.
\end{itemize} 

The rest of the paper is organized as follows.  Section \ref{sec:Works} illustrates the current state of the art for the privacy preserving record linkage problem. In Section \ref{sec:Def}, we introduce some basic definitions and the privacy model adopted in our solution. Section \ref{sec:App} provides an overview of our approach, while in Section \ref{sec:Min} and Section \ref{sec:Emb} we describe the major components in our schema. The experiment results are reported in Section \ref{sec:Exp}. Finally, we conclude the paper in Section \ref{sec:Conc}

\section{Related Works}\label{sec:Works}

In this section, we present an overview of the techniques on privacy preserving record linkage.  We carefully illustrate the secure mapping mechanism proposed by Scannepieco et al. \cite{Scannapieco:2007} that represents the closest technique to our approach.   

\subsection*{Private Record Linkage Techniques}

There is a broad variety of strategies proposed by the scientific community to tackle the private record linkage problem, which differ in privacy notions, protocol models, and the type of objects required to be matched. In the following, we distinguish these techniques  in three major categories: \textit{secure transformations}, \textit{Secure Multiparty Computation}, and \textit{hybrid methods}.

{\bf Secure transformation}   techniques aim to perform the linkage of the records after some transformations have been applied to the original data. The typical scenario involves three parties, where two parties have the data, and using secure transformation techniques, they sent the data to a third party whose task is to perform the matching. In this framework, two major strategies have been proposed: hashing and  embedding. Approaches based on hashing functions  try to match  strings by hashing the original data and computing similarity measures after the hash functions are applied. Although, these techniques are quite popular, they do not provide a formal bound on the distance in the hashed space. In addition, they are subject to dictionary attack, and if the hash function is disclosed the entire security and privacy are compromised.  Examples of hashing techniques are Bloom filter \cite{Bloom}, $q$-grams hashing \cite{Blind}, and TFIDF Hashing \cite{Al-Lawati:2005}. 

A more recent example of the embedding strategy for record linkage is the approach proposed by Scannepieco et. al.  \cite{Scannapieco:2007} that uses SparseMap \cite{Hjaltason} to embed  strings into a vector space and perform  matching in this new space.
The core of this approach relies on the Lipschitz embedding \cite{Bourgain,Johnson} which consists in projecting each original data point $s$ into a new space using a base $\mathcal{B}$ of subsets $A_i$, $\mathcal{B}=\{A_1,A_2,\dots,A_k\}$, so that each point $s$ is mapped via the embedding function $\rho$ into a vector $\bar{s}\in\mathbb{R}^k$ , where each coordinate $\bar{s}_i=\min_{x\in A_i}\{d_{Edit}(x,s)\}$, for $i=1,2,\dots,k$. The distance between vectors in the embedded space is measured by using the Euclidean distance $d'$, while the original space the distance metric is the Edit distance $d_{Edit}$.
The general approach guarantees the privacy by showing that no information about the datasets are disclosed, since only a base of random string is shared between the parties. However, to improve the performance of this approach the authors in \cite{Scannapieco:2007} proposed several heuristics that aim to better select the random strings in the base. Although these techniques improve the performance, they may disclose sensitive information. Indeed, as the protocol is designed, the shared base is optimized according the data of one party. Then, if the other party is malicious, it may potentially break the privacy by inferring the original from the structure of the base.

A key feature of embedding approaches is that the distance in the original space can be put in relationship with the distance in the new space depending on the distortion induced by the embedding map. Unfortunately, general embedding function are computational expensive to apply and some heuristic are needed. Moreover, bounding the distortion induced by some embedding may be technically challenging. For this reason, in this paper we present a embedding technique based on the formal model of differential privacy.

{\bf Secure Multiparty Computation}  (SMC) techniques cast the record linkage problem into a  secure communication framework. In this scenario, several parties are involved in the protocol where the communication is done using cryptography.
The key idea is that the computation itself should reveal no more than whatever may be revealed by examining the input and output of each party. An important theoretical result  in the cryptographic area \cite{Yao:1982} shows that any computational functions can be computed in this setting. Motivate by this, several works have been proposed in the literature. For example, when the exact match is considered the record linkage problem can  be interpreted as a set intersection problem \cite{Yao:86}. To mention, the  work in \cite{crypto:2008} investigates the SMC approach in privacy preserving data mining. While in principle the private record linkage problem can be solved using SMC and cryptography, the computational and communication cost of these methods turn out  too great in real application.

{\bf Hybrid} methods combine anonymization or secure transformation techniques with SMC techniques with the aim of reducing SMC cost. Inan et al.\cite{Inan:2008} proposed a composed strategy based on SMC and sanitization to achieve a trade off between privacy and utility. This work has been further extended in \cite{Inan:2010} by differentially private blocking followed by SMC techniques for matching record pairs in matched blocks instead of matching all record pairs. On the framework of blocking record linkage, the work in \cite{EvangelistaCSM10} combines machine learning techniques in defining the blocking functions, showing interesting results in the utility.
Hybrid techniques provide a good trade-off between privacy and accuracy, the SMC step still involves high computational cost and the impact of the blocking on the linkage accuracy is not clearly understood. 
\section{Preliminaries} \label{sec:Def}

In this section, we introduce some notations and definitions related to our approach. First, we briefly present  notions concerning strings records and the concept of embedding. Second, we review the model of differential privacy which is used as our privacy model. An overview of the frequent symbols used in the paper is summarized in Table \ref{tab:symb}.

\subsection{Basic Definitions}

Let $\Sigma$ be a finite alphabet, we denote by $x=x_0x_1\cdots x_{n-1}$ a string of length $n$ where each symbol $x_i$ is defined in $\Sigma$. Moreover, we denote by $|x|$ the length of the string $x$. 
Given a string $x$ of length $n$, we represent a substring from position $i$ to $j$ in $x$ with $x[i,j]=x_ix_{i+1}\cdots x_{j}$, where $0\le i\le j\le n-1$.  
A common similarity measure between strings is the Edit distance, known as Levenshtein distance, which measures the number of edit operations needed to transform a string into the other one. 
\begin{definition}[Edit Distance \cite{levenshtein1966}]
The Edit distance between two strings $x$ and $y$ is defined as the minimum number of character edit operations necessary to transform $x$ into $y$. A single character edit operation can either replace, delete or insert a character in $x$ or in $y$. We denote the Edit distance between $x$ and $y$ by $d_{Edit}(x, y)$.
\end{definition}
Using this definition of the similarity metric, we can denote by $(\Sigma^*;d_{Edit})$ the metric space formed by the pair: space of all possible strings, and Edit distance. 

In this scenario, we informally introduce the notion of embedding used in the paper as the map $\rho:\Sigma^*\rightarrow\mathbb{R}^k$, where $k$ is the dimensionality of the embedded space. Where the distance in the new space is the Euclidean distance.

\begin{table}
\centering
\caption{Table of frequent symbols}\label{tab:symb}
\scriptsize
\begin{tabular}{|c|l|} \hline
{\bf Symbol}&{\bf Description}\\ \hline
$D_A,D_B$ & Databases for party $A$ and party $B$\\ \hline
$\epsilon$ & Private parameter\\ \hline
$k$ & Size of the base\\ \hline
$N$ & Size of the databases\\ \hline
$\mathcal{B}$ & Shared base for the embedding\\ \hline
$x,y,z$ & Strings\\ \hline
$\bar{x},\bar{y},\bar{z}$ & Vectors\\ \hline
$\alpha, \beta, \omega$ & Substrings\\ \hline
$|\cdot|$ & Length for strings, substring, grams\\ \hline
$d_{Edit}(\cdot)$ & Edit distance\\ \hline
$d'(\cdot)$ & Distance in the embedded space\\ \hline
$ed$ & Threshold in terms of Edit operations\\ \hline
$th$ & Threshold in terms of $d'$\\ 
\hline\end{tabular}
\vspace{-0.2 cm}
\end{table}

\subsection{Differential Privacy}

Differential privacy \cite{Dwork06} is a recent notion of privacy that aims to protect the disclosure of information when statistical data are released. The differential privacy mechanism guarantees that the computation returned by a randomized algorithm is insensitive to change in any particular individual record of the data in input.
\begin{definition}[Differential Privacy \cite{McSherry2006}]
A non interactive privacy mechanism $M$ has $\epsilon$-differential privacy if for any two
input sets (databases) $D_A$ and $D_B$ with symmetric difference one (neighbor databases), and for any set of outcomes $S\subseteq Range(M)$,
\begin{eqnarray}
Pr[M(D_A) \in S]\le exp(\epsilon) \times Pr[M(D_B) \in S] .
\end{eqnarray}
\end{definition}
where $\epsilon$ is the privacy parameter (also referred to as privacy budget). Intuitively, lower value of $\epsilon$ implies stronger privacy guarantees, and vice versa.

This mechanism has two important properties that are extensively used when differential privacy computations are combined. These two properties are known as \textit{sequential} and \textit{parallel} compositions \cite{McSherry:2009}.
The former states that any sequence of computations that each provides differential privacy in isolation also provides differential privacy in sequence.
\begin{theorem}[Sequential Composition \cite{McSherry:2009}]
Let $M_i$ be a non-interactive privacy mechanism which provides $\epsilon_i$-differential privacy. Then, a sequence of $M_i(D)$ over the database $D$ provides $(\sum_{i}\epsilon_i)$-differential privacy.
\end{theorem}
The latter instead holds when the computations involved are disjoint. In this case, the privacy cost does not accumulate but depends only on the worst guarantee.
\begin{theorem}[Parallel Composition \cite{McSherry:2009}]
Let $M_i$ be a non-interactive privacy mechanism which provides $\epsilon_i$-differential privacy. Then, a sequence of disjoint $M_i(D)$ over the database $D$ provides $(\max_{i}\epsilon_i)$-differential privacy.
\end{theorem}

In literature, there are two major techniques to achieve differential privacy: \textit{Laplace noise} \cite{McSherry2006} and \textit{Exponential Mechanism} \cite{McSherry2007}. Both of these strategies are base on the concept of \textit{global sensitivity} \cite{McSherry2006} for the function to compute. 
\begin{definition}[Global Sensitivity \cite{McSherry2006}] 
For any two neighbor databases $D_A$ and $D_B$, the global sensitivity for any function $f:D\rightarrow\mathbb{R}^n$ is defined as:
\begin{eqnarray}
GS(f) := \max_{D_A,D_B}\|f(D_A)-f(D_B)\|_1.
\end{eqnarray}
\end{definition}

In the rest of the paper we restrict our attention on counting queries, which can be proved to have global sensitivity $GS(count)=1$. The Laplace mechanism is used in the following sections of the paper to construct differentially private algorithms, so for this we briefly discuss this mechanism. Let  $f$ be a function, and  $\epsilon$ be the privacy parameter, then by adding noise to the result $f(D)$ we obtain a differential privacy mechanism. The noise is generated from  a Laplace distribution with probability density function $pdf(x|\lambda)=\frac{1}{2\lambda}e^{-|x|/\lambda}$, where the parameter $\lambda$ is determinate by $\epsilon$ and $GS(f)$. 
\begin{theorem}[Laplace Mechanism \cite{McSherry2006}]
For any function $f:D\rightarrow\mathbb{R}^n$, the mechanism $M(D)$ that returns:
\begin{eqnarray}
M(D)=f(D)+Lap(GS(f)/\epsilon),
\end{eqnarray}
guarantees $\epsilon$-differential privacy.
\end{theorem}

\begin{figure}[t]
	\centering
		\includegraphics[scale = 0.65]{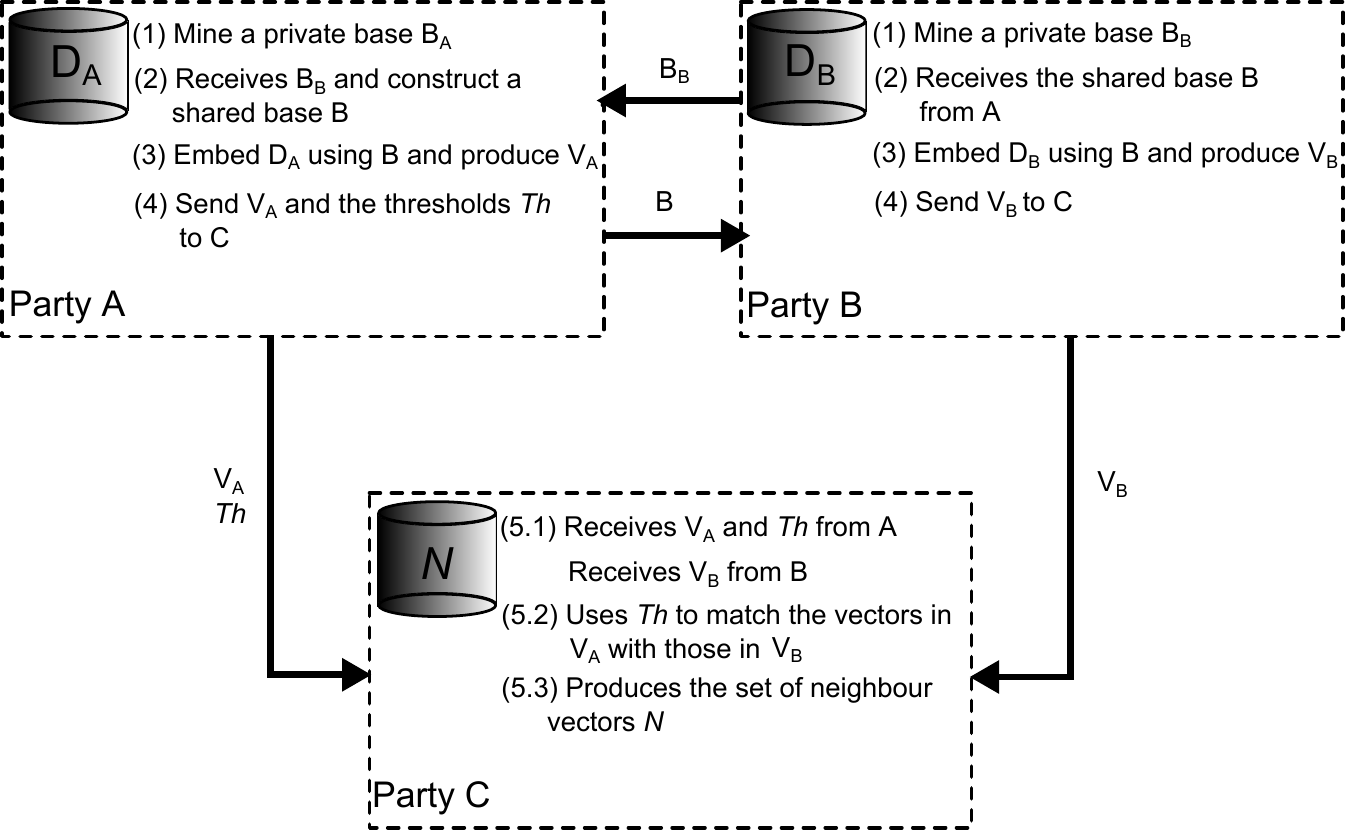}
	\caption{{\small Overview of the Secure Protocol.}}
	\label{fig:prot}
	\vspace{-0.2 cm}
\end{figure}

\section{ Overview of our Solution } \label{sec:App}

 In this section we introduce our matching mechanism that relies on the idea of matching the string records in a metric space
obtained by embedding the original datasets. We propose an embedding technique based on \textit{grams}, and a matching protocol to match close records (i.e. within a fixed number of edit operations). The base is produced by mining the original data by using \textit{differential privacy} mining algorithms.

\begin{figure}[b]
	\centering
	\subfigure[{\small {\bf Mining and Base Generation} }]{
		\includegraphics[scale=0.3]{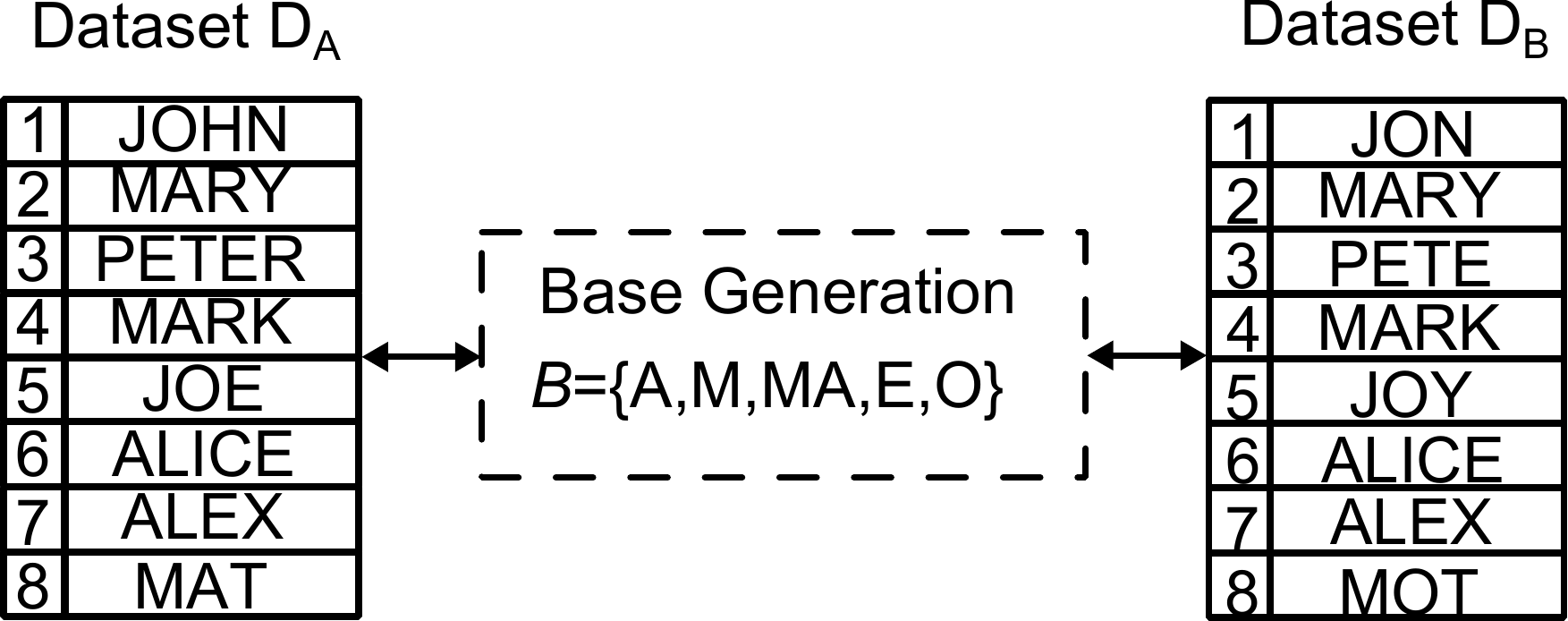}
		\label{fig:ORData}
        }\hspace{0.5cm}
	\subfigure[{\small {\bf Embedded Data}}]{
		\includegraphics[scale=0.3]{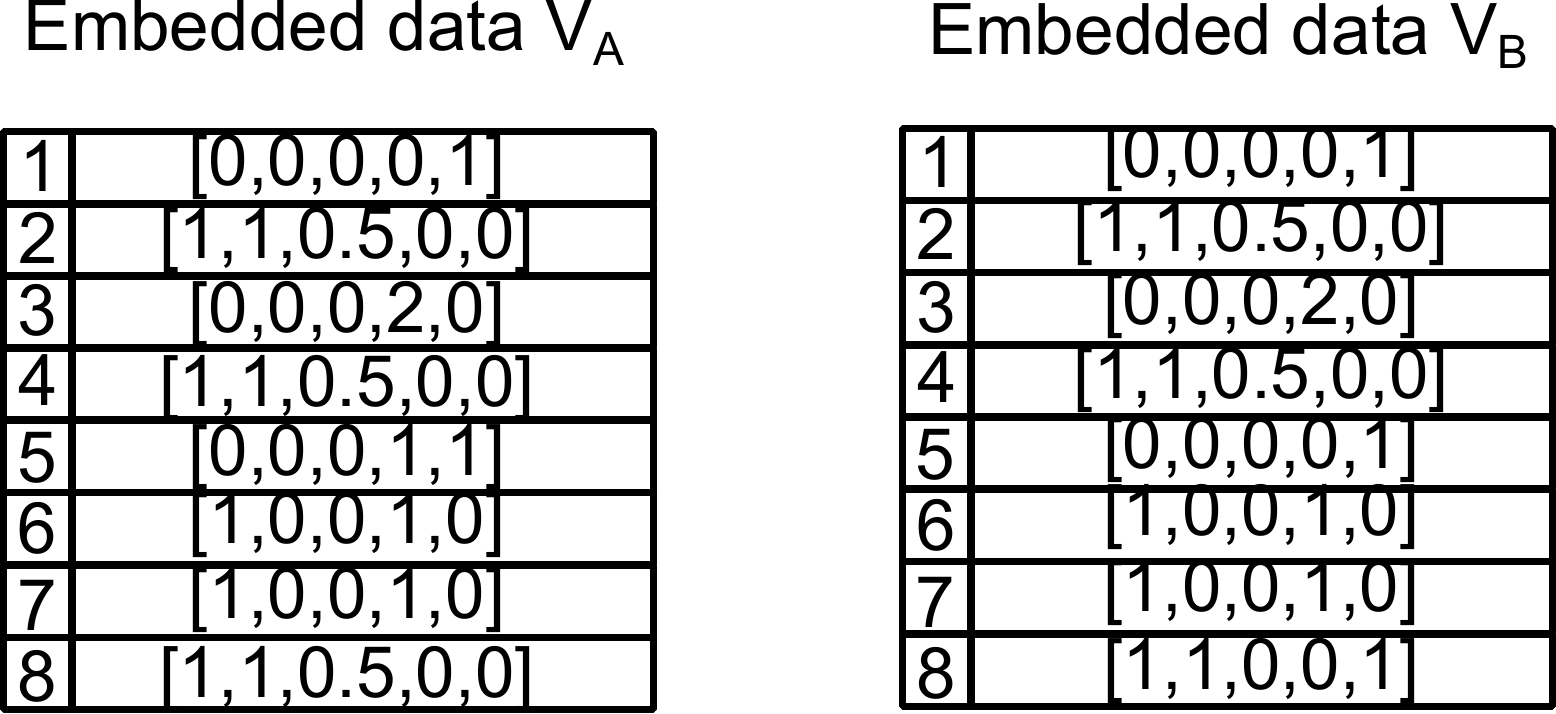}
		\label{fig:EmData}
	}
	\caption{{\small Example of Mining, Base generation and Embedding of the data.}}
	\label{fig:example}
	\vspace{-0.2cm}
\end{figure}

Our first contribution is an embedding strategy which maps the original data space into a vector space by projecting each string in the databases on a base formed by a set of frequent \emph{grams}. A gram of length $q$ is a substring $x_0x_1\cdots x_{q-1}$ with $q$ symbols.  Each party starts to build a base for the embedding by mining grams from the strings in its own database, this phase is denoted as {\bf mining phase}. This process is performed with a guarantee of differential privacy, so that the parties involved in the protocol can share their bases and determinate a common base for the embedding without disclosing any sensitive information of individual records.
When a final base is determined, each party embeds its data using the common base, and the matching is performed in the embedded space. We denote this step as {\bf embedding phase}. Our overall protocol is illustrated in Figure \ref{fig:prot}. Our strategy requires the presence of a third trusted party denoted by $C$, whose task consists in matching the records in the embedded space. A summary of the  steps is reported as follows.
\begin{enumerate}
\item {\bf Mining Phase:} Parties $A$ and $B$ apply a differentially private algorithm to mine their respective databases $D_A$ and $D_B$, and compute \textit{private} bases $\mathcal{B}_A$ and $\mathcal{B}_B$.

\item {\bf Base Generation:} One of the two parties is in charge of merging the two bases and producing a shared base $\mathcal{B}$ of frequent variable length grams. 

\item {\bf Embedding Phase:} Each party $A$ and $B$, by using the shared base, embeds its own data and generates a set of vectors $V_A$ and $V_B$ respectively, representing the strings in the original datasets.  These sets are sent to the third party $C$.

\item {\bf Threshold Generation:} Given a maximum value of edit distance $ed$ in the original space, the party $A$ generates a set of threshold values $Th=\{th_1, th_2, \dots, th_N\}$ for the distance in the embedding space for each of its string $s_i$, $i=1,2,\dots,N$ in its own dataset $D_A$ to use in the matching phase. Each $th_i$ is a personalized threshold for the string $s_i$ and it represents the maximum distance between $\bar{s}$ and $\bar{z}$, such that the string $z$ is close to $s_i$. This set of thresholds is sent to $C$.

\item {\bf Matching Phase:} The third party $C$, for each vector $\bar{s}_i\in V_A$ returns a set of neighbor vectors $\mathcal{N}(\bar{s}_i)$ computed as follows:
\begin{eqnarray}
\mathcal{N}(\bar{s}_i):=\{y\in V_B\textrm{ s. t. } d'(\bar{s}_i,\bar{y})\le th_i\}.
\end{eqnarray}

\end{enumerate}
Figure \ref{fig:ORData} illustrated the mining and base generation phase. The party $A$ and $B$ mine their respective datasets and produce a share private base formed by the following grams $\{$\texttt{A,M,MA,E,O}$\}$. This shared base is used by each party to embed their own data and produce the set of vectors as in Figure \ref{fig:EmData}. In the following sections, we describe each phase of our approach in details.

\section{Mining phase} \label{sec:Min}

Contrary to the SparseMap approach, we  construct a base \textit{mined} from the original data for mainly two reasons. First, we take advantage of the fact that usually in record linkage scenarios the strings that are being matched have similar properties (e.g. same alphabet, similar length, etc. ), so by constructing a base from the original dataset we can capture this information. Second, a base randomly generated cannot well represent every datasets since it is defined in a generic way and not data dependent.  

Our idea consists in forming a base to embed the strings in the original space by mining the \textit{frequent} grams in the database. Formally, given a positive integer $k$, a minimum length $q_{min}$ and a maximum length $q_{max}$, we are interested in mining the top-$k$ frequents $q$-grams where $q\in[q_{min},q_{max}]$, and use this set to construct the base for the embedding. Intuitively, in this way we can obtain a base set that is a good representative for all the strings in the databases. This idea has been successfully applied in the approximate string matching framework \cite{Yang08cost,Li:2007}. The problem of mining sequential patterns poses several challenges, and the literature is rich of efficient solutions to address this problem \cite{Han:2000,PrefixSpan}. However, a direct use of the mined base from a party in the protocol may disclose information about the strings in its database. For example, Atzori et al. \cite{AtzoriBGP05}  investigates the problem of disclosing sensitive information in frequent itemsets mining using blocking anonymity. In our approach, we consider two approaches to guarantee that the base produced from the mining phase satisfies the differential privacy notion. In the following, we present two techniques to mine the frequent grams. The first approach is an adaptation of the score perturbation \textit{frequent pattern mining} technique proposed by R. Bhaskar et al. \cite{Bhaskar:2010}, while the second one is adapted from the \textit{prefix tree} mining algorithm introduced in \cite{CFDS12kdd}. 

The motivation to propose two mining algorithms is based on the fact that the algorithm in  \cite{Bhaskar:2010} requires prior knowledge of $q_{min}$, $q_{max}$ and $k$ to guarantee the differential privacy of the grams. Clearly, if these values change or are unspecified, the algorithm has to use additional privacy budget to adjust the released data. On the other hand, the prefix tree miner approach proposed in this section does not have this limitation. Indeed, building the  prefix tree can be done only one time and does not depend on these parameters (beside for the hybrid strategy). 
Once the prefix tree is produced, it can be mined many times for frequent grams of different lengths and $k$. Therefore, this approach is preferable when some of these parameters is not specified, and an exploration of the parameter values is required. In this section we prove that our mining phase produces a base of grams that guarantees $\epsilon$-differential privacy. In the case that the two databases may be correlated as they may contain records belonging to the same owner, the total privacy parameter is split among the two parties holding the dataset, so that the overall privacy level is $\epsilon$.

\subsection{FPM Miner}

The score perturbation frequent pattern mining (FPM) proposed in \cite{Bhaskar:2010} requires in input the value $k$ and the length of the patterns $l$ to mine. As a output, it produces a list of the top-$k$ most frequent patterns of length $l$ in the dataset and this algorithm guarantees differential privacy. The original  approach runs a non-private mining algorithm to retrieve the list of frequent itemsets up to a given length $l$ first, and then given $k$ it computes the frequency of the $k$-th most frequent item $f_k$. Using this information, the frequencies of the items are truncated, and later perturbed with Laplace noise.  After a sampling process the list of top-$k$ items are reported.

In order to mine only patterns instead of items, we run the FPM by mining the frequent patterns with the non private algorithm in the first step. Then the perturbation and the truncation of the frequencies are considered only on the universe of patterns. Since, the FPM mines fixed length patterns, given the overall privacy parameter $\epsilon$, for our mining phase we can adapt this approach by running $\Delta_q=q_{max}-q_{min}+1$ runs of the FPM algorithm, where in each run the length of the pattern is $q$ for $q=q_{min},q_{min}+1,\dots,q_{max}$, the size of the set is $k$ and with privacy parameter $\epsilon/\Delta_q$. In this way, we mine the top-$k$ most frequent patterns for each length, and finally we select the top-$k$ most frequent out of them. 

\subsubsection{Privacy Analysis}

The privacy analysis follows directly from the application of the sequential composition lemma and from the result in \cite{Bhaskar:2010}. 

\begin{theorem}[FPM Miner differential privacy]
The FPM Miner guarantees $\epsilon$-differential privacy.
\begin{proof}
The FPM Miner runs $\Delta_q$ calls of FPM algorithm which has been proven to be $\epsilon/\Delta_q$-differential private in  \cite{Bhaskar:2010}. Therefore, by the \textit{sequential composition lemma} for differential privacy the overall strategy guarantees $\epsilon$-differential privacy. 
\end{proof}
\end{theorem}

\subsubsection{Complexity Analysis}

The running time for the FPM-miner follows directly from the time complexity for FPM algorithm. The authors in \cite{Bhaskar:2010} shown that the complexity for their approach is related to the non-private algorithm used to initially mine the items and compute their truncated frequencies. By running a non private miner that computes the occurrences for all the grams of length $q$,  this phase results into a $O(k_{q}N)$ factor, where $k$ is the number of $q$-grams which frequency is grater than $f_k-\gamma$, and $\gamma$ is the utility parameter.  From \cite{Bhaskar:2010}, each run of FPM requires  $O(k_{q}+k\log k_{q}+k_{q}N)$ time, hence the overall complexity of the FPM-miner is $O(\Delta_q(k'+k\log k'+k'N))$, where $k'=\max_{q\in[q_{min},q_{max}]}k_q$.

\begin{figure*}[htbp]
	\centering
	\includegraphics[scale=0.6]{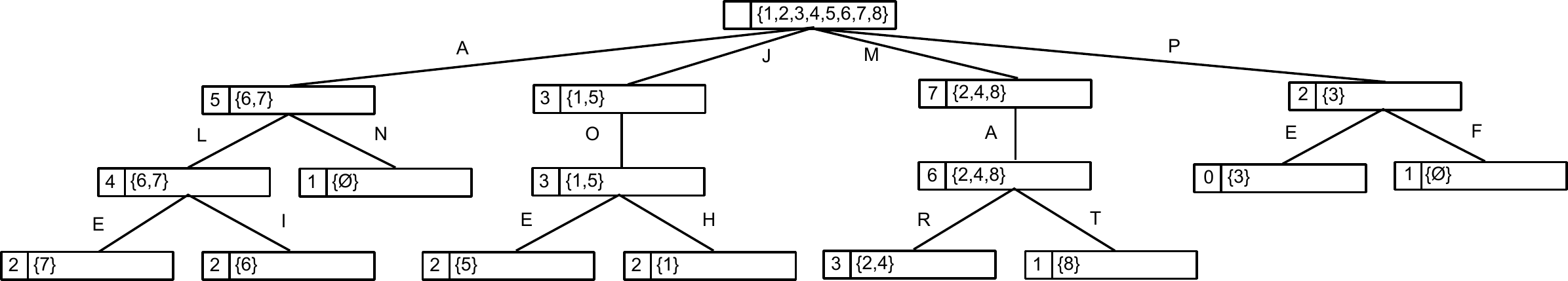}
   	\caption{{\small Prefix Tree Example from $D_A$ in Figure \ref{fig:ORData}. Each nodes has a noisy count, and the set of identifiers for the strings in the partition. Each branch of the tree is labeled with the symbol used to extend the prefix.}}
	\label{fig:PT}
	\vspace{-0.2 cm}
\end{figure*}

\subsection{Prefix-tree Miner}

In addition to the previous strategy, we adapt the mining algorithm developed to mine sequential data in \cite{CFDS12kdd}. For our mining task, we partition the space of all the possible grams using a top-down approach, where each partition is identified by a node in a prefix tree $\mathcal{T}$. Each node has the following information: a prefix $\omega$, an accumulated privacy budget and the subset of all the strings in the original database having $\omega$ as a prefix, called the partition represented by the node. Then the prefix tree is traversed and all the frequent variable length grams are reported.

\begin{algorithm}[!b]
\caption{Private Prefix-Tree Partitioner} \label{algo:PPT}
\scriptsize
\begin{algorithmic}[1]
\Procedure {PPT Part}{$node,\mathcal{T}$}
\Statex
{\bf Input:} node $node$;  private prefix-tree $\mathcal{T}$
\Statex
{\bf Output:} $\mathcal{T}$ private Prefix-Tree
\Statex
\While{(($node.h\le h_{MAX}$) \&\& ($node.budget<\epsilon$))}
\For {(every symbol $a$ in the alphabet $\Sigma$)}
\State $\omega \gets $ (path from $r$ to node) + $a$
\State $P\gets\{x\in node.set \textrm{ s. t. $\omega$ is a prefix of $x$}\}$
\State $\tilde{\epsilon} \gets \Call{ALLOCATE Budget}{node}$
\State $count\gets |P|+Lap(1/\tilde{\epsilon})$
\If{($count>\theta$)} \Comment{Non empty node}
\State Add a new node $cur$  as child of $node$, such that:
\State $cur.set\gets P$, $cur.epsilon\gets\tilde{\epsilon}$
\State $cur.label\gets \omega$,
\State $cur.budget\gets cur.budget + cur.epsilon$
\State $cur.h\gets node.h + 1$
\State \Call{PPT Part}{$cur$,$\mathcal{T}$} \Comment{Recursive call}
\EndIf
\EndFor
\EndWhile
\State \textbf{return} $\mathcal{T}$
\EndProcedure
\end{algorithmic}
\end{algorithm}

The construction of the prefix tree can be summarized as follows. Starting from the root of the tree, labeled with the empty string and representing the entire space, the database is partitioned  by extending the prefix of the current node  using the strategy in Algorithm \ref{algo:PPT}.  For every symbol $a$ in the alphabet $\Sigma$ a new node is added to the tree if the string $\omega a$ is a frequent prefix in the dataset, where $\omega$ is the string formed by the concatenation of the symbols following the path from the root to the parent of the current node. To determinate if a prefix is frequent, in lines 5 to 8 in Algorithm \ref{algo:PPT}, a counting query is issued on the partition of dataset represented by the current node and the real count is perturbed by Laplace noise to guarantee the differential privacy. This noisy count is compared against a threshold $\theta$ representing the noisy count for empty partitions. In this process,  only partitions with frequent prefixes are further refined, and when their frequencies are not large enough or a specified depth of the tree is reached the partitioning strategy stops. The allocation of the budget at each level in the tree is performed at line 6 in Algorithm \ref{algo:PPT}. In our approach, we propose several strategies to allocate the private budget in the tree: \textit{linear allocation}, \textit{exponential allocation}, \textit{adaptive}, and \textit{hybrid}. Details about these strategies are reported at the end of this section.

An example of the prefix tree obtained from the dataset $D_A$ in Figure \ref{fig:ORData} is reported in Figure \ref{fig:PT}. Each node represents a partition on the string in $D_A$, with associated the noisy count of the prefix obtained by following the path from the root to the node. For example, the prefix \texttt{AL} induces a partition formed by the strings with IDs 6 and 7 in $D_A$, and the noisy count for this prefix is 5.

After we partition the data space using the prefix-tree, we traverse the prefix tree and for each root-to-leaf path we apply the following consistency constraints as in \cite{CFDS12kdd}:
\begin{definition}[Consistency Constraints  \cite{CFDS12kdd}] 
For any node $v$ in the prefix tree, the following holds:
\begin{enumerate}
\item $count(v)\ge count(u)$, where $u$ is a child of $v$.
\item $\displaystyle count(v)\ge\sum_{u \textrm{ is a child of $v$}}count(u)$.
\end{enumerate}
\end{definition}
 
Once the consistency constraints are enforced,  we report a list of grams by traversing the tree. First of all we can notice that the frequency count for any gram $x$ is given by the sum of the frequencies of the prefixes that have $x$ as a suffix.
Therefore, for any frequent prefix $\omega$ in the tree we increase the frequency of the substring $x$ of length $q\in[q_{min},q_{max}]$ that is a suffix of $\omega=\alpha x$ by the count of $\omega$.
To speedup this step, we can adopt a more efficient construction of the prefix tree that uses prefix links that can be used to efficiently traverse the tree and extract the grams \cite{Cheng:2010} . As a final result, from this list we return the top-$k$ grams sorted by their noisy frequencies.

\subsubsection{Budget Allocation Strategies}\label{sec:All}

In this section, we describe the strategies used in our prefix tree partitioning algorithm to allocate the budget in issuing the counting query for each node. In the following we denote by $\epsilon$ the level of privacy that we aim to guarantee with our tree and by  $h_{MAX}$ the maximum depth of the prefix tree.

\begin{itemize}
\item {\bf Linear:} At each level in the tree we allocate the same amount of budget $\epsilon_0=\epsilon/h_{MAX}$ for each node. 
\item {\bf Exponential:} At the level $i$ in the tree, a node can use an amount of budget that is double of the amount spent by its parent, $\epsilon_i = 2\epsilon_{i-1}$, $i=1,\dots,h_{MAX}-1$. Since on the overall path root-leaf we need to guarantee the $\epsilon$-privacy, we start with $\epsilon_0=\frac{\epsilon}{2^{h_{MAX}+1}-1}$
\item {\bf Adaptive:} This strategy is an adaptation of the previous exponential allocation schema, where the entire remaining budget on the path is spent on the next counting query if the current node represents a non frequent prefix. 
\item {\bf Hybrid:} This strategy as the previous one, uses a threshold to decide if we use all the remaining budget  on the next counting query, and in addition the total budget is distributed in the tree according to the maximum length $q_{max}$.  In particular, we  reserve half of the total budget to the nodes on the first $q_{max}$ levels of the tree, where for each node the budget is allocated proportionally with its level in the tree. For the nodes with level greater than $q_{max}$, the remaining part is allocated in an exponential fashion. For a node at level $i$ we use the budget $\epsilon_i$ defined as follows:
\begin{eqnarray}
\epsilon_i:=\left\{\begin{array}{ll}
\frac{\epsilon(i+1)}{q_{max}(q_{max}+1)} & \textrm{if $0<i\le q_{max}$}\\
\frac{\epsilon2^{i-q_{max}-1}}{2(2^{h_{MAX}-q_{max}}-1)} & \textrm{otherwise}\\
\end{array}\right.
\end{eqnarray}
The intuition behind this strategy consists in avoiding premature pruning in the tree. Indeed, with this approach we reserve half of the total budget on the first $q_{max}$ levels of the tree, so that frequent prefixes are likely extended at least  to $q_{max}$ symbols.  This allows us to capture more occurrences of the grams present in the frequent prefixes.
\end{itemize}

\subsubsection{Privacy Analysis}

What is left to show is that the set of grams report satisfies the differential privacy. We can notice that, all the partitions produced by Algorithm \ref{algo:PPT}  on the same level of the prefix tree are disjoint since they correspond to strings with different prefixes. Hence by the \textit{parallel composition lemma} for differential privacy the overall privacy level is related to the maximum value of the private budget spend in over all the paths from the root to the leaves of the tree. Given a sequence of nodes of the tree, the overall privacy level is given by \textit{sequential composition lemma} and it can be computed by considering the sum of the privacy parameters used for each counting query in each node.

\begin{theorem}[PrefixTree differential privacy]
The Prefix tree Miner guarantees $\epsilon$-differential privacy.
\begin{proof}
The proof follows directly from the differential privacy result proposed in \cite{CFDS12kdd} since it is easy that  all the allocation strategies uses at most $\epsilon$ budget on the path root-leaf in the tree.
\end{proof}
\end{theorem}

\subsubsection{Complexity Analysis}

Algorithm \ref{algo:PPT} has running time proportional to the number of nodes in the prefix tree $\mathcal{T}$. First, we can notice that on the level $i$, we have at most all the possible prefixes of length $i$ defined on the alphabet $\Sigma$. Hence, the total number of nodes at level $i$ is bounded by $O(|\Sigma|^i)$. Its real value can be smaller due to the fact that some non frequent prefixes are not further extended. Moreover, each node performs a counting query on the partition of its parent that in the worst case requires a linear scan. Therefore each level $i$ requires $O(N|\Sigma|^i)$ operations, where $N$ is the size of the database in input. Since, in the tree there are at most $h_{MAX}$ level, the overall running time for Algorithm \ref{algo:PPT} is $O(N|\Sigma|^{h_{MAX}+1})$. In our implementation to speed up the space partitioning, we process the children of an internal node in lexicographic order with respect to their label. So that, each counting query for a child is issued on a subset of positions that have not occurrences of its previous siblings. This can be performed efficiently by storing for each node a sorted list of the positions where the prefix represented by the node occurs. Moreover, the list is maintained sorted at each parturition step without any additional computation cost.  
After the tree is constructed,  the consistency constraints requires $O(h_{MAX}^2N)$ operations as shown in  \cite{CFDS12kdd}. Finally, the running time for traversing the prefix tree is linear with the number of internal nodes in the tree that is $O(N|\Sigma|^{h_{MAX}+1})$. Therefore, the overall prefix tree mining algorithm runs in $O(N|\Sigma|^{h_{MAX}+1})$.

\section{Embedding phase}\label{sec:Emb}

In this section we describe the embedding procedure to map strings into vectors. Let  $\mathcal{B}=\{g_1,g_2,\dots,g_k\}$ be a base of $k$ grams, each string $s$ in the database is mapped into a vector $\bar{s}$ in $\mathbb{R}^k$, where each component $\bar{s}_i$ represents  the number of occurrences of the gram $g_i$ in $s$, normalized by the length of $g_i$. Let $Occ_s(g)$ denote the set of positions in $s$ where the gram $g$ occurs,
\[
Occ_s(g):=\{ i \in [0, |s|-|g|] : s[i, i+|g|-1] = g\}
\]
Each coordinate of the vector representing $s$ is defined as $\bar{s}_i = |Occ_s(g_i)|/|g_i|$, for $i=1,\dots,k$. In this new space the distance between vectors is computed using the Euclidean distance as follows:
\begin{eqnarray}
d'(\bar{x},\bar{y})=\|\bar{x}-\bar{y}\|_2=\left(\sum_{i=1}^k(\bar{x}_i-\bar{y}_i)^2\right)^{1/2}
\label{eq:dist}
\end{eqnarray}
This new distance measure can be interpreted as a weighted version of the $q$-gram distance proposed by Ukkonen \cite{Ukkonen:1992}, where the contribution of each gram is reduced by a factor equal to the length of the gram, and where the grams considered are a subset of the shared grams between the strings in the original space. Intuitively, with this mapping procedure, strings that are close in the original space remain close also in the vector space.

\subsection{Threshold Computation}\label{sec:pers}

After all the records are embedded into the new space the approximate matching between the records is performed. In particular, we are interested in matching strings that in the original space are within $ed$ edit operations. In the new space, this task is casted into the problem of finding all the vectors whose Euclidean distance is within a threshold value $th$. This threshold value plays a central role on the overall performance of the matching schema, since it decides how many vectors become candidate records in representing close strings. Therefore, it is crucial to compute a threshold value as tight as possible to the real value. 
This problem is generally very hard, since proving formal guarantees  requires  analysis on the distance distortion and properties of the embedding strategy used. For example, for the Lipschitz embedding it can be shown that the distortion for this mapping is $O(\log n)$ where $n$ is the size of the original data space \cite{Hjaltason,Bourgain}. \\
{\bf Global threshold:} This strategy aims to define a \textit{global threshold} value that can be used in matching all the records in the new space. This can be done by estimating how the original distance is distorted after the embedding map is applied. In this direction, we proposed the following upper bound for our embedding.
\begin{proposition}[Upper bound]
Given $x$ and $y$, two strings in the original space with Edit distance $d_{Edit}(x,y)\le ed$, an upper bound on $d'(\bar{x},\bar{y})$ is given by $\Delta_q\cdot ed$, where $\Delta_q=q_{max}-q_{min}+1$
\begin{proof} 
For a fixed length $q$ of the grams, and for any $ed$ value of edit distance between $x$ and $y$, the number of grams of length $q$ that can contribute  is at most $q\cdot ed$ (since on a position $i$ at most $q$ grams of length $q$ are overlapping). Since,  the base $\mathcal{B}$ used in  the embedding is a subset of all possible $q$-grams, with $q\in[q_{min},q_{max}]$ therefore it follows that:
\begin{eqnarray}
d'(\bar{x},\bar{y})=\|\bar{x}-\bar{y}\|_2\le \Delta_q\cdot ed
\end{eqnarray}
\end{proof}
\end{proposition}
This upper bound is not very tight, and does not take into account the fact that each string shares different number of grams from the base used for the embedding. For example, by referring to the scenario in Figure \ref{fig:example}, if $ed=1$ the upper bound for a global threshold in matching the embedded data is $(2-1+1)\cdot 1=2$.\\
{\bf Personalized threshold:} In contrast, in our approach we define a \textit{personalized threshold} for each string that has to be matched. This choice is motivated by the following reasons.  First, each string shares a different number of grams with the base, and for those strings that have a large number of shared grams a personalized threshold can better represent the original distance. Second, by restricting the attention in computing a threshold for each string we overcome the problem of estimating a threshold suitable for all the strings required to be matched. To compute this personalized threshold,  we use a dynamic programming algorithm which for each string computes the maximum distance in the embedded space where a vector representing a string within $ed$ edit operations can be founded. In particular, we adopt the algorithm proposed in \cite{Yang08cost}, where in our case we keep track of the impact of each gram on the final distance rather than counting only their occurrences.

Given a string $s$, we start to initialize two data structures as follows:
\begin{eqnarray}
D[i] = \sum_{g\in\mathcal{B}}{\left(\frac{\delta_s(i,g)}{|g|}\right)^2}, \quad i=0,1,\dots,|s|-1\nonumber \\
\delta_s(i,g):=\left\{\begin{array}{ll}
1 & \textrm{if $g$ overlaps $s$ at position $i$}\\
0 & \textrm{otherwise}\\
\end{array}\right.\\
P[i]=\max_{g\in\mathcal{B}}\{j<i-|g|-1 \textrm{ where } j\in Occ_s(g)\}
\end{eqnarray}
For each position $i$ in $s$, the entry $D[i]$ represents the total contribution the grams overlapping at position $i$ may have in the coordinate of $\bar{s}$ if an edit operation occurs at position $i$. On the other hand, $P[i]$ denotes the nearest position $j<i$ where a gram of the base occurs in $s$. Using these structures and given a maximum value of edit operations allowed $ed$, we are interested in computing for each string $s$ the maximum allowable distance $th$ in the embedding space. Given $i$ edit operation, the dynamic programming strategy considers two cases. First, an edit  operation does not occur at position $j$, so the contribution remain the same as in the previous position $T[i][j-1]$. Second, a edit operation occurred in $j$ then we need to add the contribution from the grams in the current position $D[i]$ to the contribution of having $i-1$ edit operations in the positions before $j$. The overall strategy is reported in Algorithm \ref{algo:th}, and the running time for this dynamic programming algorithm is $O(ed\cdot |s|)$.

\begin{algorithm}[!t]
\caption{Dynamic Algorithm for computing $th$} \label{algo:th}
\scriptsize
\begin{algorithmic}[1]
\Procedure {Personalized Th}{$s,D[\cdot],P[\cdot],ed$}
\Statex
{\bf Input:} String $s$;  $D[\cdot]$, $P[\cdot]$, and edit distance $ed$.
\Statex
{\bf Output:} Threshold value $th$
\Statex
\For{($i=0,1,\dots,ed$)}
\For{($j = 0,1,..., |s|-1$)}
\State $T[i][j] = \max\{T[i][j-1],T[i-1][P[j]]+D[j]\}$
\EndFor
\EndFor
\State \textbf{return} $th=\sqrt{T[ed][j]}$
\EndProcedure
\end{algorithmic}
\end{algorithm}

\subsection{Protocol Complexity Analysis}\label{sec:comp}

In this section we analyzed the complexity of our protocol in terms of time and communication complexity.

The mining phase in our strategy leads to $O(k'N)$ for the FPM miner and $(N|\Sigma|^{h_{MAX}+1})$ for the Prefix tree based approach. After the base $\mathcal{B}$ is computed the databases are mapped into the vector space in the embedding phase.  Let $l$ be the maximum length of the strings in the databases, then this step requires $O(lkN)$ operations, since each gram can have at most $l$ occurrences in each string. When the strings are transformed into vectors, each pair of vectors can be matched using the Euclidean distance in $O(k)$ time. 

In addition to the complexity analysis reported above, we also investigated the communication complexity required by our protocol. 
We are interested in providing an upper bound on the overall amount of bits transmitted between the party. Without lost of generality, we can assume that every string in the database requires $O(\ln l)$ bits to be represented. Therefore, the shared base $\mathcal{B}$ formed by the mined grams is transmitted by using $O(k\ln q_{max})$ bits. On the other hand, the embedded records are vectors in $\mathbb{R}^k$, and therefore they require $O(kN)$ bits to be represented. The overall communication complexity for this protocol is $O(k(N+\ln q_{max}))$.

\begin{table}[t]
\centering
\scriptsize
\caption{Experimental Datasets Statistics}\label{tab:datasets}
\begin{tabular}{|c|c|c|c|c|} \hline
{\bf Dataset}& $N$ & $l_{max}$ & $l_{min}$ & $l_{avg}$\\ \hline
\texttt{NAMES} & 150000 & 15 & 4 & 7 \\ \hline
\texttt{CITIES} & 5000 & 23 & 3 & 8 \\
\hline
\end{tabular}
\vspace{-0.3 cm}
\end{table}

\begin{table}[t]
\centering
\caption{Table of experiment and algorithm parameters}\label{tab:para}
\scriptsize
\begin{tabular}{|c|l|c|} \hline
{\bf Symbol}&{\bf Description}&{\bf Default values}\\ \hline
$\epsilon$ & Private parameter & 0.1\\ \hline
$k$ & Size of the base & 75\\ \hline
$q_{min}$ & Min length of the grams  & 1\\ \hline 
$q_{max}$ & Max length of the grams  & 3\\ \hline 
$ed$ &  Edit operations & $\{0, 1, 2\}$\\ \hline
$h_{MAX}$ &  Max depth of prefix tree & $l_{avg}$\\ \hline
$\theta$ &  Threshold for noisy count & as in \cite{CFDS12kdd}\\ \hline
\end{tabular}
\vspace{-0.3 cm}
\end{table}

\section{Experimental Results}\label{sec:Exp}

In this section, we present experimental results obtained with our private record linkage protocol. Our goal is to understand the impact of the mining and embedding phases on the overall utility of our protocol as well as the dependency from the private parameter $\epsilon$ and the dimensionality $k$. We compare our approach with Scannapieco et. al's \cite{Scannapieco:2007} matching schema to show the improvement in scalability, the benefit of a stronger privacy model and the use of personalized threshold in the matching schema while achieving comparable data utility.

The protocols were implemented in Java, and the simulations were conducted on an Intel Core i5 2.5Ghz PC with 4GB of RAM. In the experiments, we use two real datasets \texttt{NAMES} \footnote{\texttt{NAMES} is publicly  available at United States Census (\href{http://www.census.gov/genealogy/www/data/2000surnames/}{http://www.census.gov/genealogy/www/data/2000surnames/})}, and \texttt{CITIES}. The first set contains a list of the most frequent surnames from the Census 2000 in United States. The second dataset is a list of the top 5000 most populated cities in United States in 2008.  Some of the statistics of the datasets are summarized in Table \ref{tab:datasets}, where $l_{max}$, $l_{min}$ and $l_{avg}$ are the maximal, minimal and average lengths of the strings, respectively.

Before presenting the results, we describe the simulation scenario. First of all, the datasets do not contain duplicated records, and the simulation proceeds as follows. The party $A$ holds a dataset $D_A$ while the party $B$ has a perturbed copy $D_B$ of $D_A$, where each string is corrupted up to $ed$ edit operations. So in this way we can run the experiments for different values of edit operations. Below, we first study the utility of the mined base, and then we examine the utility of the overall linkage results. We use the $F_1$ score as utility metric \cite{van1979}, which combines the precision and recall in matching the strings between the datasets.

The experiment and algorithm parameters, if not specified in the description of the simulations, assume the default value as reported in Table \ref{tab:para}. 

\subsection{Mining Utility}

In our first set of experiments, we examine the utility of the mined frequent grams with respect to the privacy parameter $\epsilon$ and the value $k$. For this set of experiments due to space limitations, we report only the results from the \texttt{NAMES} dataset. Similar results can be observed on the \texttt{CITIES} dataset.

\begin{figure}[!t]
	\centering
	\includegraphics[scale=0.2]{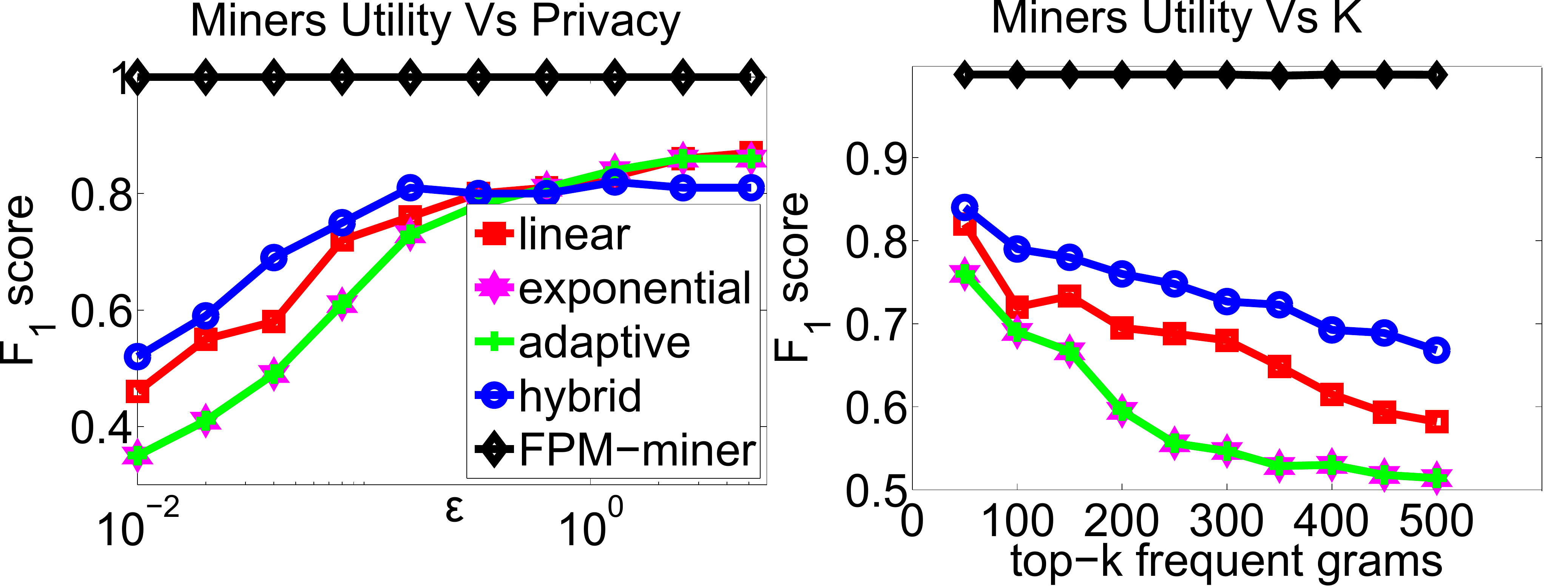}
	\vspace{-0.2cm}
	\caption{{\small Utility for mining algorithms}}
	\label{fig:UtilPTree}
	\vspace{-0.2cm}
\end{figure}

Figure \ref{fig:UtilPTree} illustrates the utility if the results produced by the FPM and prefix tree miner approaches with respect to different values of privacy level and number of grams mined. From Figure \ref{fig:UtilPTree} we can see that increasing the privacy parameter $\epsilon$ (i.e. decreasing the privacy level) leads to better performances. Among the allocation strategies proposed in Section \ref{sec:All}, the hybrid strategy provides significantly better utility for small $\epsilon$ values, but slightly worse utility when $\epsilon>0.2$. This confirms the fact that this strategy has been designed to reduce the early pruning in the tree due to small $\epsilon$ values. For the rest of the simulations on the linkage, we will run the prefix tree miner by using the linear, and hybrid allocation strategies, since they provide the best utility  for  low and high lever of privacy respectively. On the right part of Figure \ref{fig:UtilPTree}, we illustrate the dependency between the utility and the number of grams mined ($k$). As we can see, all the allocation strategies have the same behavior, where increasing the value of $k$ decreases the utility of the mined base. This result is related to the fact that the prefix tree is based on growing frequent prefixes. So when $k$ is large some of the occurrence of the grams may not appear in the tree due to the fact that prefixes are not extended enough to capture their occurrences. We can also observe that, the hybrid strategy performs consistently better than to the other strategies. 

The mining performance provided by the FPM-miner, in our simulations, outperforms that of the prefix tree miner in both the scenarios described above. Although the FPM-miner provides utility results close to the optimal value, it suffers of the fact that the values for $q_{min}$, $q_{max}$ and $k$ must be specified a priori. Indeed, if the optimal value for these parameters is unknown, or one of these parameters changes the FPM-miner must run again and it increases the total privacy budget spent. On the other hand, the prefix tree miner with linear, exponential and adaptive budget allocation strategies does not have this limitation, since the construction of prefix tree does not depend on these parameters. Only traversing the tree requires knowledge of $q_{min}$, $q_{max}$ and $k$, and it is done on data  that are already differentially private.

\begin{figure}[!t]
	\centering
	\includegraphics[scale=0.2]{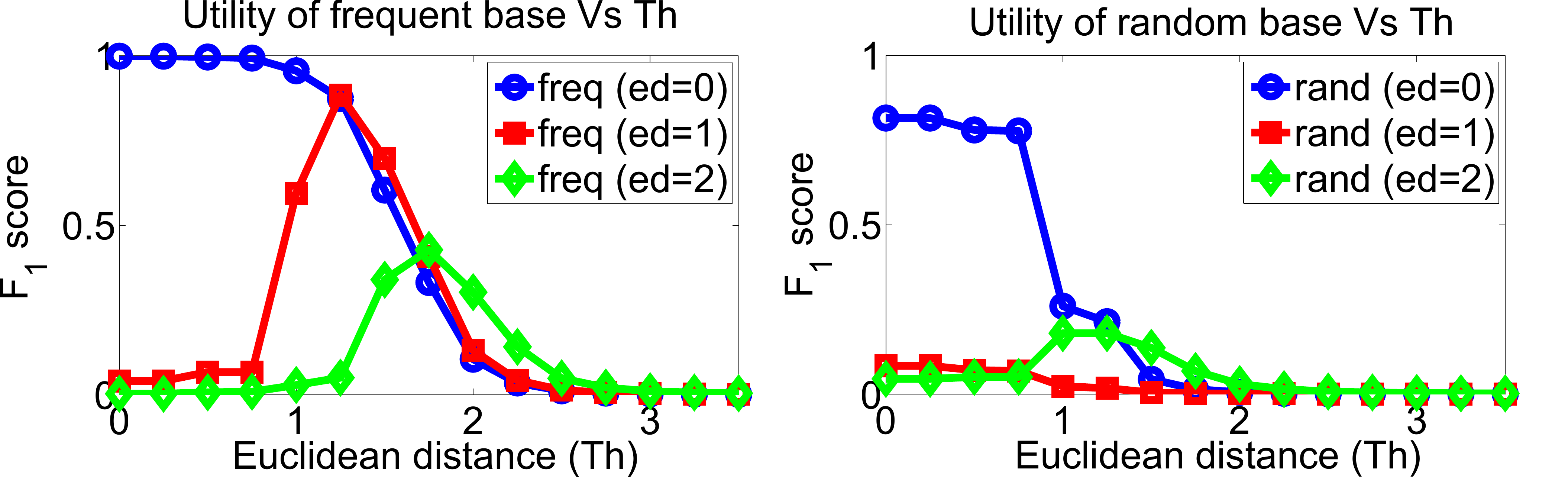}
	\vspace{-0.2cm}
   	\caption{{\small Utility and threshold: (Left) frequent grams, (Right) random grams}}
	\label{fig:UtilFreq}
	\vspace{-0.2 cm}
\end{figure}

\subsection{Linkage Utility}

In this part of the simulations, we illustrate the advantages of having a personalized vs a global threshold, the dependency of the utility results for different parameters, as well as the overall performance of our mechanism.

\subsubsection{Global Vs Personalized Threshold}

First, we study the importance of using frequent grams over random grams in the base. Figure \ref{fig:UtilFreq} reports the utility in terms of $F_1$ score for different values of global threshold ($Th$) in the embedded space, with a direct comparison between random and frequent grams. We tested the embedding strategy on the \texttt{NAMES} dataset, by allowing an approximate matching with the number of edit operations up to 2, with $k=100$. For our scenario, two edit operations represent a high level of perturbation of the original dataset, since they correspond to change roughly up to  35\% of the symbols (in average) on the strings. From the graph, it is evident that frequent grams lead to a considerable improvement in the utility with respect to random grams (an improvement from 20\% to 60\%). This result is justified by the fact that a base of frequent grams is more likely to share a higher number of grams  with the strings. Therefore,  the frequent  base can better preserve the distance and better represent the original data with respect to a random base.  In addition, we can also observe that when approximate matching is allowed the utility decreases as the  edit operations increase. Indeed, the matching becomes harder to perform since more grams are affected by the edit operations. 
 
Second, form Figure \ref{fig:UtilFreq}, we can observe that using a global threshold in the matching we have an optimal point that maximizes the utility. However, it is not easy to compute it a priori. For example, by using the upper bound obtained in Section \ref{sec:Emb} we have that $Th\le q_{max}\cdot ed=6$, and in this case is too large to provide good utility. For this reason, the rest of the results are obtained by using the personalized threshold approach.

\begin{figure}[!t]
	\centering
	\includegraphics[scale=0.2]{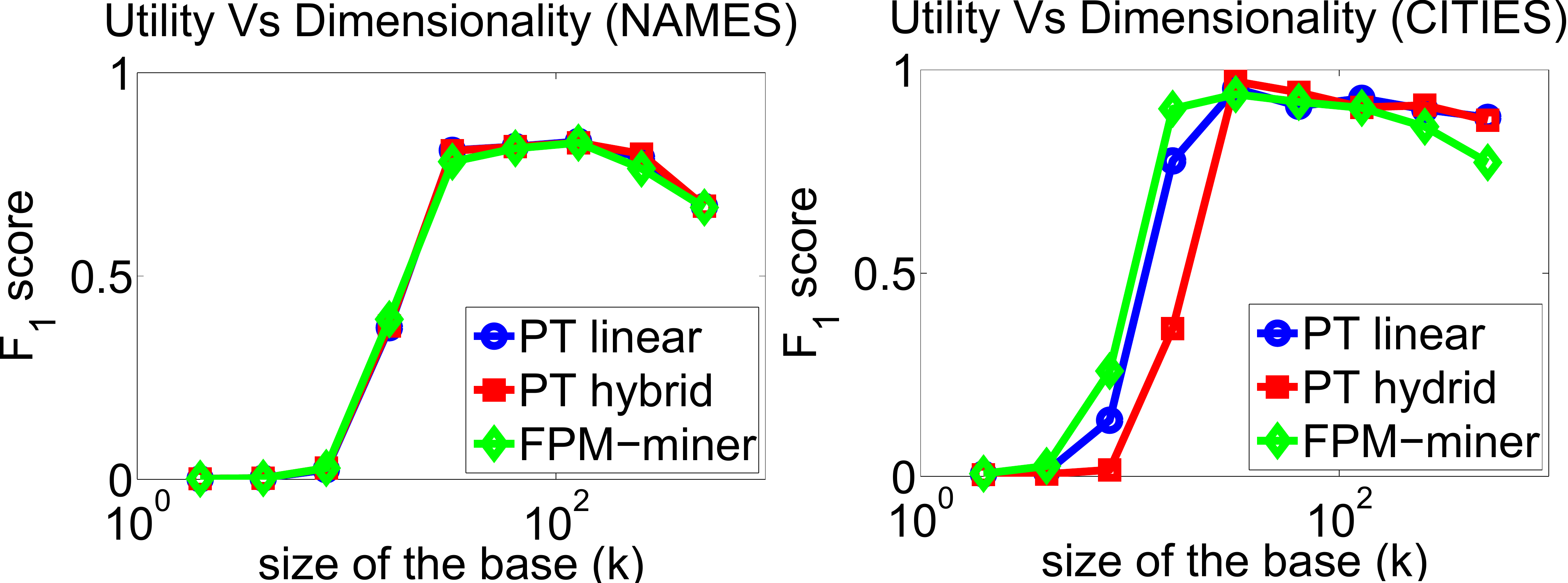}
	\caption{{\small Impact of the base size, $ed=1$}}
	\label{fig:UtilDim}
	\vspace{-0.2 cm}
\end{figure}

\subsubsection{Impact of the base size}

In Figure \ref{fig:UtilDim}, we examine the impact of the size of the shared base $k$ on the utility of the linkage results. In these simulations, we consider an approximate matching scenario with $ed=1$.
The experiments show that the utility increases by increasing the dimensionality till a maximum value is reached. We can observe that when a base has size in between 50 and 75 grams, our approach provides  better performance for both datasets considered.  After that, a further increase of the dimensionality degenerates the utility of the results.  This phenomena is very peculiar, since  usually  the dimensionality in the embedded space helps with the utility. However, it can be explained with the fact that a large base can better preserve the distance in the space, but it also increases the number of grams that potentially we can lose in case of edit operations. Therefore, the threshold value increases and the performance decreases as shown in Figure \ref{fig:UtilDim}. 

\subsubsection{Impact of the privacy parameter}

\begin{figure}[!t]
	\centering
	\subfigure[{\small {\bf Utility results with \texttt{NAMES} dataset}}]{
		\includegraphics[scale=0.2]{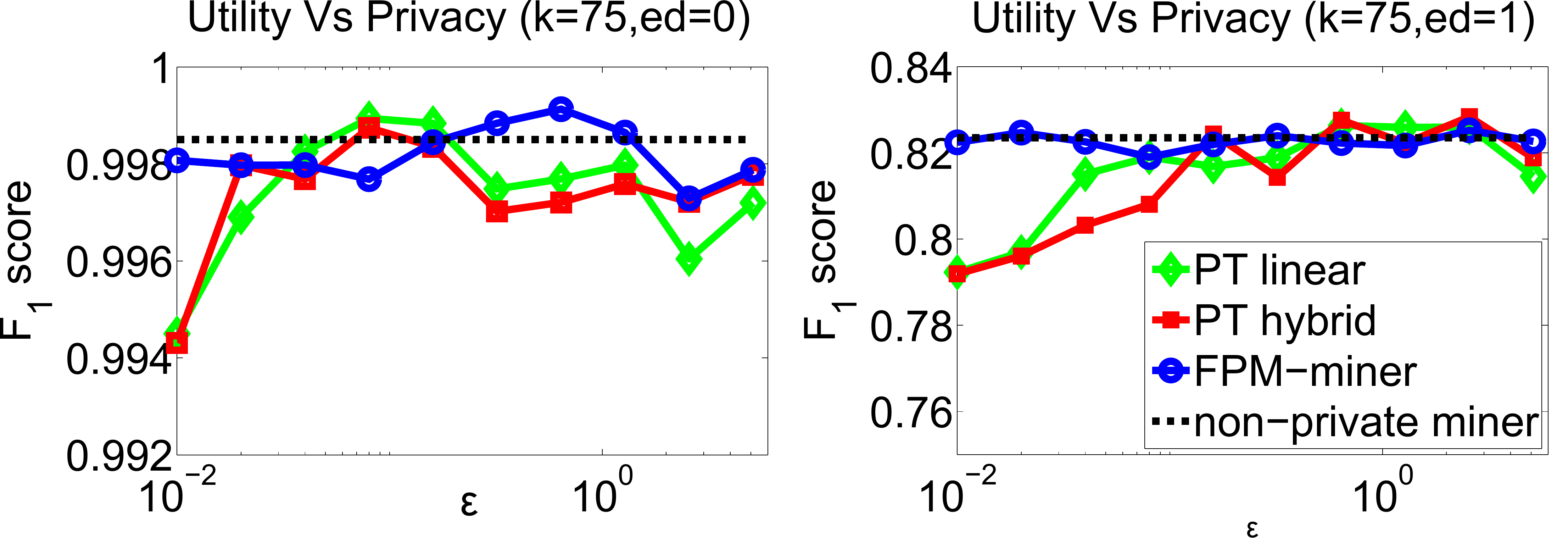}
		\label{fig:UtilPrivNames}
        }
	\subfigure[{\small {\bf Utility results with \texttt{CITIES} dataset}}]{
		\includegraphics[scale=0.2]{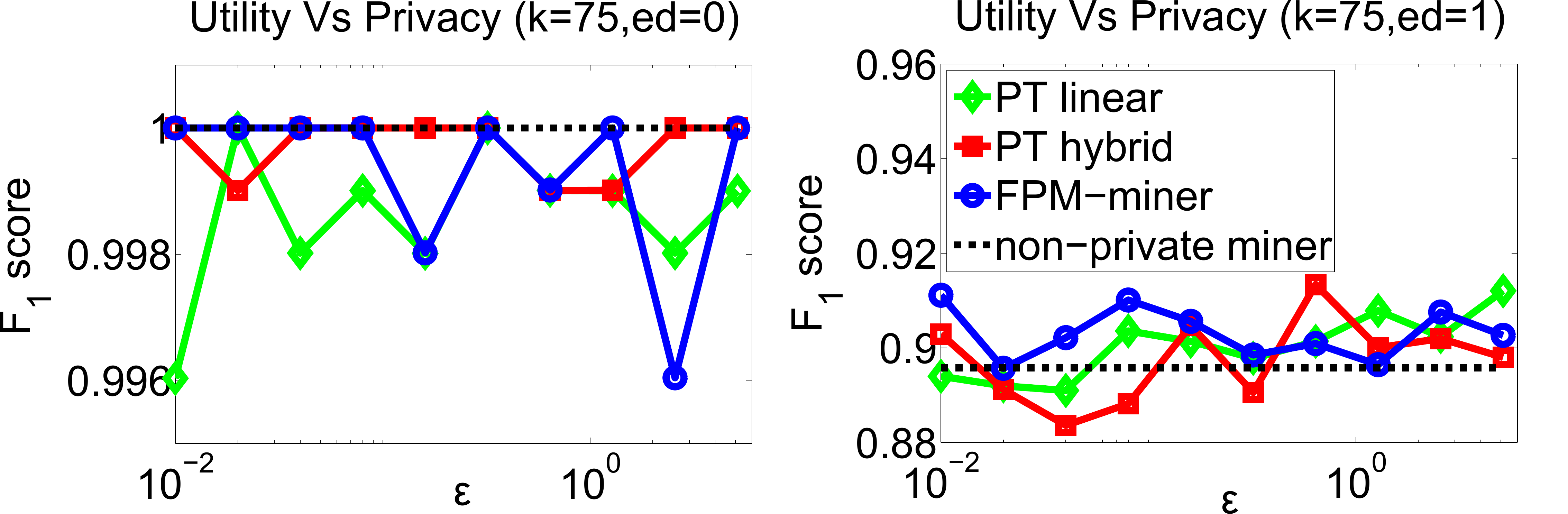}
		\label{fig:UtilPrivCities}
        }
        \caption{{\small Impact of the privacy parameter }}
        \label{fig:UtilPriv}
        \vspace{-0.2 cm}
\end{figure}

The relationship between the privacy parameter $\epsilon$ and the utility of our protocol is reported in Figure \ref{fig:UtilPriv} for the datasets considered. In these graphs, we also compare the results provided by private miners with  non private miner. As we can see, the utility of the protocols that use private miners approach the results obtained from the non private algorithm  as $\epsilon$ increases.  Moreover in Figure  \ref{fig:UtilPriv} we can see that by just allowing one edit operation the utilities of the matching results for all the approaches shifts down by at most 18\% for the datasets considered. This result points out the hardness of solving record linkage when approximate matching is allowed. In the case of exact match the utility provided by the private mechanisms are with the 0.4\% of the non private one for every $\epsilon$ value tested. While one edit is allowed, this gap slightly increases. This behavior is more evident for the mining approach based on the prefix tree.
From this set of simulations, we can infer that the embedding phase plays a crucial step in the overall performance of our algorithm. Since, as shown in Figure  \ref{fig:UtilPriv},  the final utility does not have strong dependency on the utility form the mining phase. 
\subsubsection{Impact of edit operations}

\begin{figure}[t]
	\centering
	\includegraphics[scale=0.2]{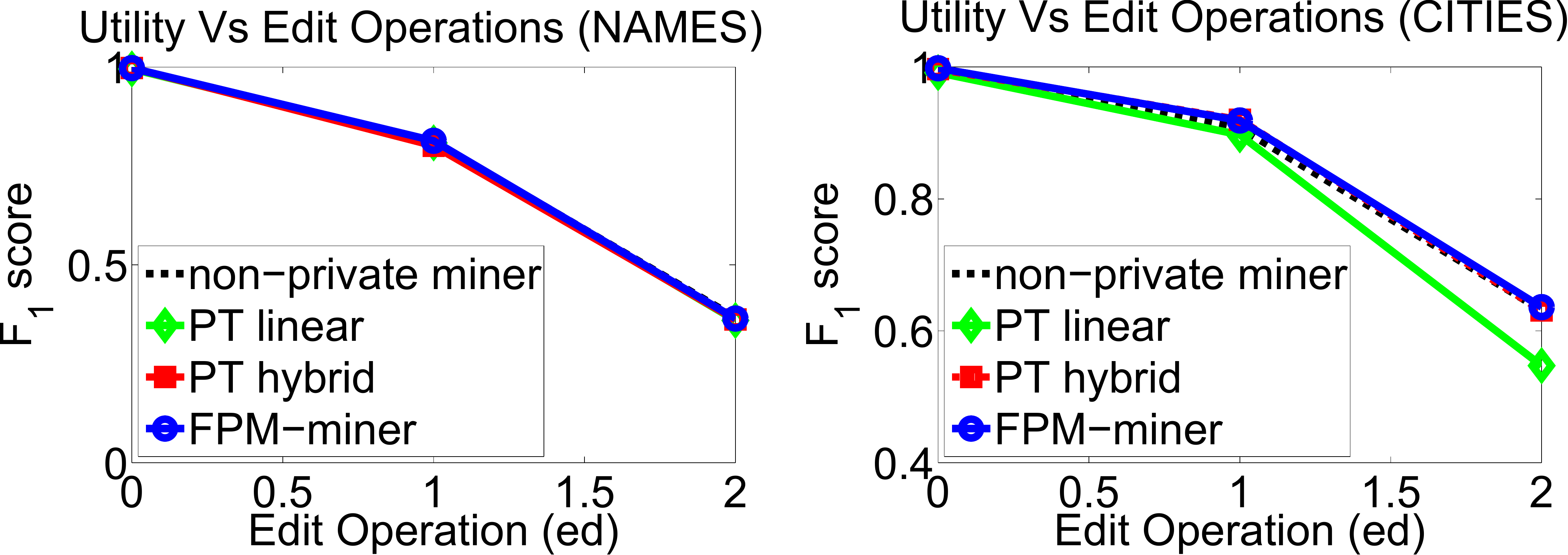}
   	\caption{{\small Impact of edit: {\bf (Left)} \texttt{NAMES},  {\bf (Right)} \texttt{CITIES}. }}
	\label{fig:UtilEdit}
	\vspace{-0.2 cm}
\end{figure}

In Figure \ref{fig:UtilEdit}, we present the impact of edit operations on the utility of the protocol.  
These pictures show that as the number of edit operations increases the overall utility decreases for all the strategies tested. The edit operations considered in these simulations vary from 0 (exact match) to 2 edits (35\% of the avg length of the string in the dataset). We can notice that the $F_1$ score decreases from 0.99 in the case of exact matching  to 0.36 when 2 edits are allowed for the \texttt{NAMES} dataset, while for the \texttt{CITIES} dataset stops at 0.60. This phenomena is due to two important aspects. First, as we mentioned earlier, when the edits  increase, more grams could be lost between strings and the base formed by the frequent grams mined in the first phase. Second, given a string $s$ the size of the set of possible matching strings within $ed$ edit operations $\mathcal{N}(s,ed)$ increases exponentially with $ed$, indeed $|\mathcal{N}(s,ed)|\le |s|^{ed}|\Sigma|^{ed}$.
Concerning the performances, we can notice that the results reported by FPM-miner and prefix tree hybrid approach are close to the utility of the results provided by the non private miner.

\begin{figure}[t]
	\centering
	\includegraphics[scale=0.2]{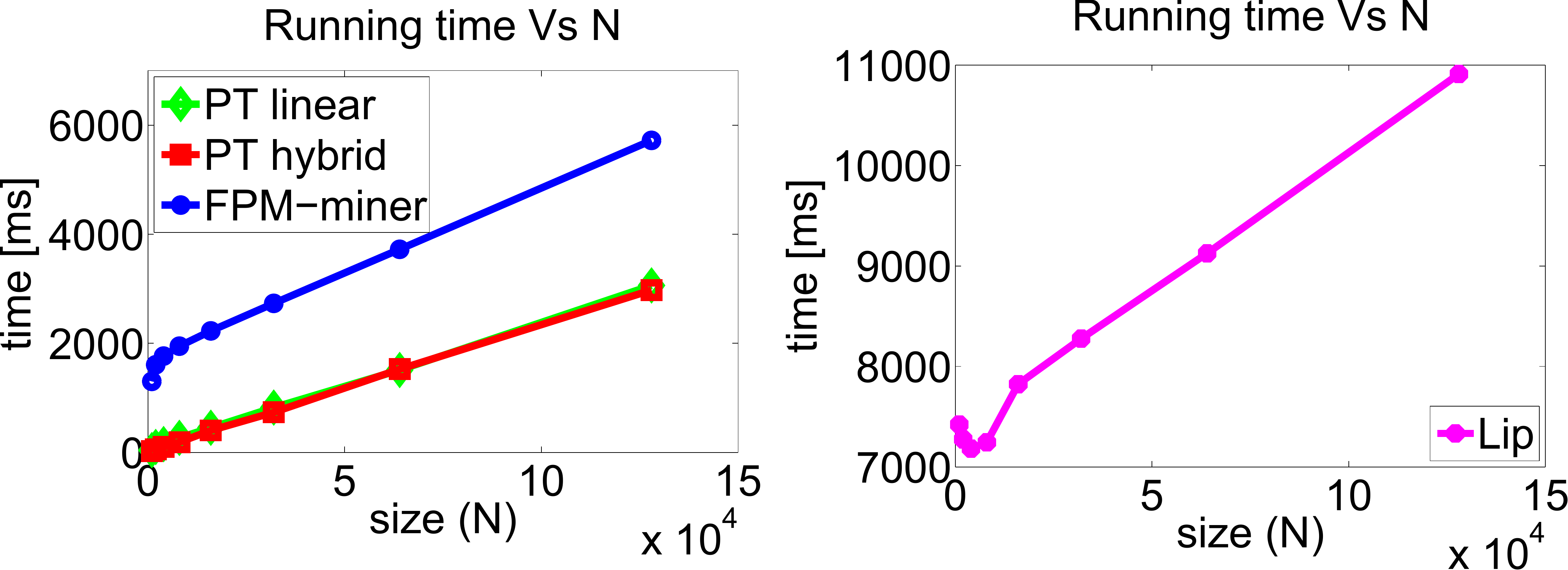}
   	\caption{{\small Running Time Vs $N$: {\bf (Left)} frequent variable length grams embedding, {\bf (Right)} Lipschitz embedding. }}
	\label{fig:TimeSize}
	\vspace{-0.2 cm}
\end{figure}

\subsubsection{Protocol Performances} 
\begin{figure}[t]
	\centering
	\centering
	\includegraphics[scale=0.2]{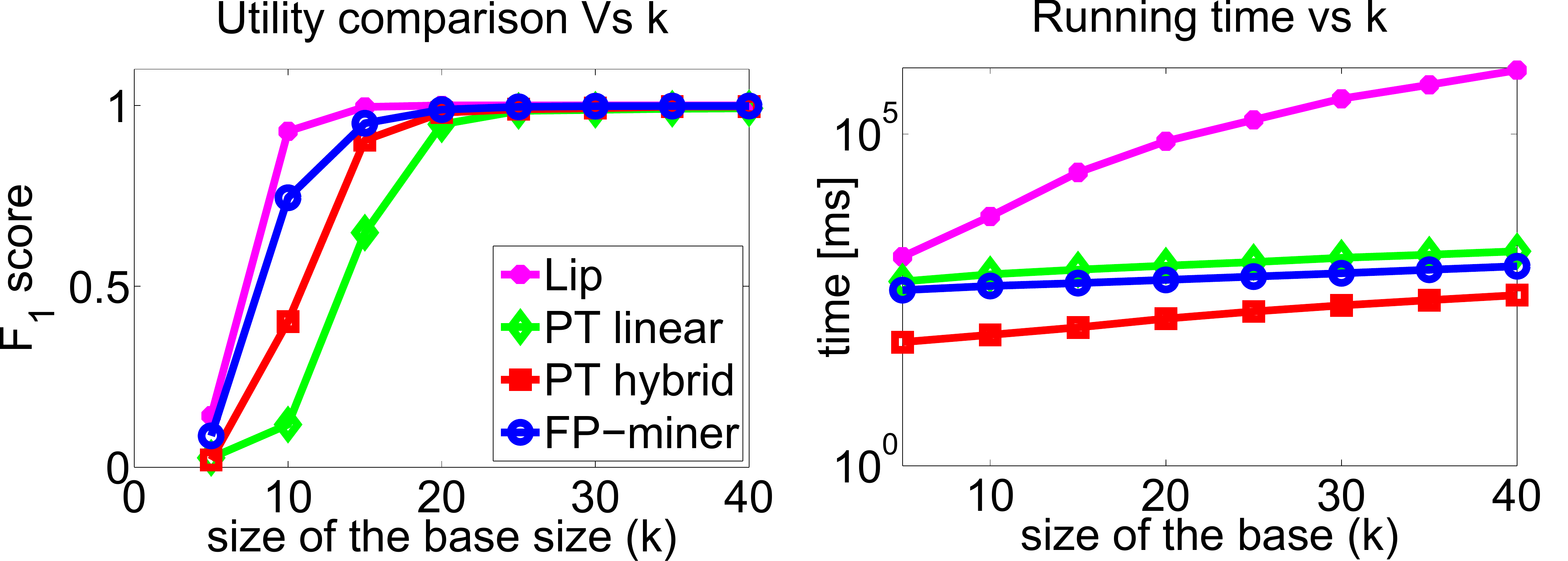}
	\caption{{\small Comparison between Lipschitz and frequent grams embedding}}
	\label{fig:Comp}
	\vspace{-0.2 cm}
\end{figure}

To examine the scalability of our strategy, we performed several experiments with different dataset sizes, and the results are reported in Figure \ref{fig:TimeSize}. The running time is measured in milliseconds [ms], and it consists in the time needed to mine the base for each party, to combine the bases to form the shared base, to embed the data, and finally to compute the personalized threshold. In addition, we consider only exact matching, since  the Lipschitz mechanism in \cite{Scannapieco:2007} requires priory knowledge of the threshold value in the embedding space to allow approximate matching.
As we can see from Figure  \ref{fig:TimeSize}, the running time for our protocol is linear with the size of the dataset considered, and this allows good scalability. These results empathizes the fact that on real data the performance of the algorithm are more realistic than the results in the complexity analysis pointed out in Section \ref{sec:comp}. Figure  \ref{fig:TimeSize} shows also the running time for the Lipschitz approach with heuristic proposed by Scannapieco et al. in \cite{Scannapieco:2007}. For this schema, we measure the time required to generate the base using the heuristic and produce the embedding map. As we can see, also this approach scales linear with the size of the dataset, but our technique requires almost half of the running time to achieve similar utility results. Indeed, for Lipschitz we choose a size of the base equals to 12 that as shown in  Figure \ref{fig:Comp} maximizes the utility and reduces the running time with respect to the dimensionality. 

In particular, for the Lipschitz approach we generate a candidate set of random strings of length 10 , and in our simulation we use a candidate set of 1000 random strings ($O(\ln N)$) as suggested in \cite{Scannapieco:2007}. In Figure \ref{fig:Comp} we test the approaches with different base sizes on the \texttt{CITIES} dataset. As we can see from the curves, Lipschitz provides a better utility for smaller base, and our strategies achieve similar results for $k>20$. Although  the secure protocol in  \cite{Scannapieco:2007} provides high level of accuracy, it requires priori knowledge of the threshold value, and this is not always possible. In addition, the dependency of the running time with respect to the dimensionality is exponential, while with our approach is only linear.

\section{Conclusions} \label{sec:Conc}

In this paper, as a global achievement we presented a novel strategy to perform privacy preserving record linkage for string records. First, we tested and adapted the differential privacy mining techniques presented in  \cite{Bhaskar:2010} and \cite{CFDS12kdd} to mine frequent variable length grams required in our embedding strategy. Second, we introduced an embedding technique to perform a secure transformation that use a private base extracted from the original data. As the final contribution, we used the concept of personalized threshold in matching records  which allows us to compute a threshold value for each record to match and does not require any priori knowledge. Our overall strategy presents comparable performance with the technique proposed by Scannapieco et al. in \cite{Scannapieco:2007}, and well fits in the secure data transformation framework  by satisfying the strong privacy model of differential privacy, and presenting good scalability.

Future works will address more complex allocation strategies for the prefix tree miner. For instance, we will consider how to take advantage of the statistical properties for the English words to issue counting queries. Moreover, we will investigate the possibility of proving formal bound on the utility of the results by assuming some statistical distributions for the datasets.

\section{Acknowledgment}
 This material is based upon work supported by the National Science Foundation under Grant No. 1117763.
%
{\scriptsize
\bibliographystyle{abbrv}
\bibliography{PPRL-EmbeddingGrams}  
}
%
%

\end{document}